\documentclass[preprint,aps,floatfix]{revtex4}
\usepackage{hyperref}
\everymath={\displaystyle}
\usepackage{young}
\usepackage{rotating}
\usepackage{longtable}
\def\1{\mbox{l\hspace{-0.53em}1}}

\usepackage{multirow}
\newlength{\AccoHaut}

\begin{document}
\title{Highly excited negative parity baryons in the $1/N_c$ expansion}

\author{N. Matagne\footnote{E-mail address: nicolas.matagne@umons.ac.be}}
\affiliation{University of Mons, Service de Physique Nucl\'eaire et 
Subnucl\'eaire,Place du Parc 20, B-7000 Mons, Belgium}

\author{Fl. Stancu\footnote{E-mail address: fstancu@ulg.ac.be}}
\affiliation{University of Li\`ege, Institute of Physics B5, Sart Tilman,
B-4000 Li\`ege 1, Belgium}

\date{\today}

\begin{abstract}
The  masses of experimentally known highly excited baryons of negative parity supposed to 
belong  to the  $[{\bf 70},\ell^-]$ multiplets  ($\ell$ = 1,2,3) of the $N = 3$ band
are calculated in the $1/N_c$ expansion  method to order $1/N_c$ 
by using a procedure  which allows to considerably reduce the number of linearly independent operators 
entering the mass formula.  The numerical fits to present data show
that the coefficients encoding the QCD dynamics have large,  comparable values, for the flavor and spin operators. It implies that these operators 
contribute dominantly to the flavor-spin SU(6) symmetry breaking, like  for the $[{\bf 70},1^-]$ multiplet of the $N = 1$ band.
\end{abstract}

\maketitle

 

\section{Introduction}

The $1/N_c$ expansion method,  where $N_c$ is the number of colors,  \cite{HOOFT,WITTEN,Gervais:1983wq,DM} 
is a powerful and systematic tool  to study ground state baryons \cite{Jenk1,DJM94,DJM95,CGO94}.
The method is based on the observation that, for $N_f$ flavors, the ground state baryons display an exact 
contracted SU(2$N_f$) symmetry when $N_c \rightarrow \infty$.  At large, but finite $N_c$, 
this symmetry is broken by contributions of order of $1/N_c$,  leading to mass splittings.  

Subsequently, efforts  have been made to extend this method to excited states. 
These states can be grouped into the so-called excitation bands $N = 1, 2, 3, {\it etc.}$ following a harmonic oscillator
notation.  In this way, one can organize the states into SU(6) $\times$ O(3) multiplets. 
So far the resonances corresponding to the $N = 1$ band  have drawn a particular
attention, being well known experimentally.  It turned out that the problem is more complicated 
technically than for the ground states, because these states belong to the SU(6) $\times$ O(3)   $[{\bf 70},1^-]$ multiplet,
thus have mixed symmetric orbital and flavor-spin parts of the total wave function.  
In such a case the   SU(2$N_f$)  symmetry is broken at order $\mathcal{O}(N^0_c)$.
The standard analysis is based  on the separation of the system into a ground
state core + an excited quark, either for $N_f = 2$ 
\cite{CGKM,Goi97,PY1,PY2,CCGL,CaCa98,Pirjol:2003ye,Cohen:2003tb} or for $N_f = 3$ \cite{SGS}. 
A simpler method, avoiding this separation has been proposed in Ref. \cite{Matagne:2006dj}
for $N_f = 2$ and  extended to  $N_f = 3$ in Refs.  \cite{Matagne:2008kb,Matagne:2011fr}.

The  $N = 2$ band contains five  SU(6) $\times$ O(3) multiplets from which four have a physical relevance.  
The  $[{\bf 56'},0^+]$  and  $[{\bf 56},2^+]$, having  symmetric orbital 
and spin-flavor states,  have been analysed in Refs. \cite{Carlson:2000zr} and \cite{GSS03} respectively, in close analogy
to  ground states. The masses of the multiplet of mixed orbital symmetry $[\bf 70, \ell^+]$,  with $\ell$ = 0 and 2, 
have been calculated by extending the ground state + excited quark method to an excited symmetric core + excited quark   \cite{Matagne:2005gd}
for $N_f = 2$. The method has been extended to   $N_f = 3$ in Ref. \cite{Matagne:2006zf}.

The $N = 3$ band contains eight  SU(6) $\times$ O(3) multiplets. In the notation of Ref. \cite{Stancu:1991cz} 
these are $[{\bf 56},1^-]$,  $[{\bf 56},3^-]$, $[{\bf 70'},1^-]$,  $[{\bf 70''},1^-]$, $[{\bf 70},2^-]$, $[{\bf 70},3^-]$, $[{\bf 20},1^-]$ and $[{\bf 20},3^-]$,
where  $[{\bf 70'},1^-]$ and $[{\bf 70''},1^-]$ correspond to  radial excitations.  
This classification provides 45 non-strange states ( 1 state $N9/2^-$, 1 state $\Delta 9/2^-$,  5 states $N7/2^-$, 2 states $\Delta 7/2^-$, 8 states $N5/2^-$,  4 states 
$\Delta 5/2^-$, 9 states $N3/2^-$, 5 states $\Delta 3/2^-$, 7 states $N1/2^-$ and 3 states $\Delta 1/2^-$). 
On the other hand 
in the 1900 MeV - 2400 MeV region only about ten non-strange resonances have been observed so far.
The interest in the $N = 3$ band has been largely hindered from a theoretical analysis in the $1/N_c$ expansion 
method,  because of the scarcity of experimental data on the one hand,  and because of its complex multiplet structure on the other hand.  
To our knowledge the only $1/N_c$ expansion study existing so far is that of Ref.   \cite{Goity:2007sc} in conjunction with Regge trajectories.
The analysis included a series of  multiplets   belonging to the $N = 1, 2, 3, 4, 5$ and 6 bands, in particular the $[{\bf 70},3^-]$ multiplet of $N = 3$.
The mass operator was reduced to a few terms containing simplified operators, considered to capture the main features of
the spectrum, but the only term of interest was   the  leading spin-flavor singlet term proportional to $N_c$.

The $N = 4$ band has 17 SU(6) $\times$ O(3) multiplets \cite{Stassart:1997vk} from which only the $[{\bf 56},4^+]$ has been
analysed in the $1/N_c$ expansion method, being the lowest one and also technically simple,  due to its symmetric  orbital and spin-flavor parts
\cite{Matagne:2004pm}.

The above studies revealed a systematic dependence of the contribution of the dominant terms in the mass formula,
with the excitation energy, or alternatively with the band number, as presented in Ref. \cite{Matagne:2005gd}. It turns out 
that the coefficient $c_1$ 
of the leading spin-flavor singlet term, proportional to $N_c$,  is raising linearly with $N$. It was also found that the coefficient of the spin-orbit  operator
having matrix elements of order $\mathcal{O}(N^0_c)$, decreases with $N$ and tends to vanish at large excitation energy. The coefficient of the spin-spin term,  with matrix elements of order $\mathcal{O}(N^{-1}_c)$,  which brings the largest contribution to the splitting,  decreases with the
excitation energy.  Results for the $N =3$ band were however absent in that analysis.  Note that the energy dependence of the mass formula
obtained in the $1/N_c$ expansion method  is remarkably compatible with the energy dependence obtained within the framework
of quark models with a chromomagnetic hyperfine interaction \cite{Semay:2007cv}.

Here we present the first  systematic attempt  towards  studying   the $N = 3$  band in the $1/N_c$ expansion method. 
We include all mixed symmetric multiplets  $[{\bf 70}, \ell^-]$ ($\ell = 1, 2$ and 3) of the band,   therefore more experimental data to
analyze.   An ultimate aim is to 
see whether or not  the results are compatible with the systematic analysis of Ref. \cite{Matagne:2005gd} and to clarify the role of
the isospin operator, found so important in the $N = 1$ band. 

At this stage it is useful to mention that both the symmetric core + excited quark procedure 
and our way of handling the problem  of mixed symmetric states are algebraic methods 
in the spirit of the Gell-Mann-Okubo mass formula. There is no radial dependence in the picture. 
The symmetric core + excited quark was originally proposed \cite{Goi97} as an extension of the 
ground state treatment to excited states and was inspired by the Hartree picture. 
 In this way, in the flavor-spin space,  the problem was 
reduced to the knowledge of matrix elements of the SU(2N$_f$) generators between symmetric 
states, already known from the ground state studies. Accordingly, the wave function was approximately 
given by the coupling of an excited quark to a ground state core of $N_c-1$ quarks, 
without performing antisymmetrisation. 
In our approach,  all identical quarks are treated on the same footing and we have an exact wave 
function in the orbital-flavor-spin space. As a price,  the knowledge of the matrix elements of the  
SU(2N$_f$) generators between mixed symmetric states is required. We have calculated and provided 
all matrix elements for SU(4) in Ref.  \cite{Matagne:2006dj}  and for SU(6) in 
Refs.  \cite{Matagne:2008kb,Matagne:2011fr}, by considering an extension of the Wigner-Eckart 
theorem. While in the symmetric core + excited quark procedure the number of terms entering the mass 
formula is excessively large our mass formula has less terms and is physically  more transparent. 
The calculated spectrum pointed out the important role of the isospin operator (indirectly of the flavor-spin 
operator, as a part of the  SU(2N$_f$) Casimir operator),  systematically neglected in the symmetric 
core + an excited quark procedure. We consider that a basic aim of the $1/N_c$ expansion  
is to find the most dominant terms with a physical meaning.

Later on, the symmetric core + an excited quark approach was strongly supported by the authors of 
Ref. \cite{Pirjol:2007ed}. Starting from a general large $N_c$ constituent quark model  Hamiltonian 
and an exact wave function,  they used transformation properties of states and interactions under the 
permutation group $S_{N_c}$ and arrived at  an expectation value of the Hamiltonian in terms of 
matrix elements  of the approximate wave function of Ref. \cite{CCGL} 
and operators acting either on the symmetric core or on the excited quark. The difference 
with the algebraic methods is that the space degree of freedom enter the discussion. As a consequence 
extra terms appear  in the mass formula. They were gathered together to match various quark models 
Hamiltonians. This matching implies constraints on the ratios of different coefficients (expressed in 
terms of radial integrals).  The numerical application to the Isgur-Karl model was partially successful 
\cite{Galeta:2009pn} by reproducing the spin-spin term but the tensor term could not be reproduced. 
More work and a deeper understanding is therefore desirable.   


Before describing the $1/N_c$ expansion method let us recall some elements of the history 
of the $N = 3$ band within the framework of the constituent 
quark model.  An important wave of interest has been trigerred by the need of finding an
assignment of the $D_{35} (1930)$ resonance announced by Cutkosky et al. \cite{Cutkosky:1976av}.
The quark model calculations of  Cutkosky and Hendrick \cite{Cutkosky:1977kd} and later on of Capstick and Isgur 
\cite{Capstick:1986bm}, incorporating a linear confinement and relativistic effects, predicted  a mass of about 200 MeV above the experimental
value.  An earlier analysis based on  sum rules derived in a harmonic oscillator basis  by  Dalitz et al. \cite{Dalitz1977} provided 
a mass of 2088 $\pm$ 25 MeV 
describing  this resonance as a  a pure $[{\bf 56},1^-]$, $J = 5/2$ state  of the  $N = 3$ band,  following 
the suggestion of Cutkosky et al. \cite{Cutkosky:1976av}.  A similar  mass range has been obtained  in Ref. 
\cite{Stancu:1991cz} in a semirelativistic constituent quark model with a linear confinement and a chromomagnetic 
interaction.  
The spin independent part of the model used in Ref. \cite{Stancu:1991cz},
which has a linear confinement,  makes the  $[{\bf 56},1^-]$
multiplet the lowest one among those compatible  with the quantum numbers of
the $D_{35} (1930)$ resonance. However this resonance remains an open problem in quark models, inasmuch as its mass
is about 200 MeV  too high above the experimental value.

As already mentioned, here we are concerned with resonances which can be interpreted as members of the mixed symmetric multiplets
of the $N = 3$ band.  An important incentive to this work  was that there is new experimental interest, for example,  in the
photo-production of $\eta$ mesons off protons which suggest a  new resonance
$N(2070) D_{15}$ which can belong to the $N = 3$ band \cite{Bartholomy:2007}.
Moreover a recent multichannel partial wave analysis 
including high lying resonances, in the so-called fourth resonance region \cite{Anisovich:2011ye,Anisovich:2011fc}. 
suggests  the existence of a high-lying spin quartet 
\begin{equation}\label{quartet}
N (2150) 3/2^-, \, N (2060) 5/2^-, \, N (2190) 7/2^-, \, N (2250) 9/2^- \, 
\end{equation}
with $L$ = 3.  In the following we shall compare this suggestion with our predictions. 

The paper is organized as follows. In the next section we introduce the mass operator defined within the $1/N_c$ expansion
method.  In Section III we present the results of four distinct fits of the dynamical coefficients in the mass formula and 
calculate the mass of the fitted resonances obtained from one of these numerical fits.  We discuss  our results in a general context 
by analogy to results obtained for the $N$ = 1 band and compare our interpretation of resonances with that of 
Refs. \cite{Bartholomy:2007,Anisovich:2011ye,Anisovich:2011fc}. 
Some conclusions are drawn in the last section.

\section{The mass operator}

\begin{table*}[h!]
\caption{Operators and their coefficients in the mass formula obtained from 
numerical fits. The values of $c_i$ and $d_i$ are indicated under the heading Fit $n\ (n=1,2,3,4)$.}
\label{operators}{\scriptsize
\renewcommand{\arraystretch}{2} 
\begin{tabular}{lrrrr}
\hline
\hline
Operator \hspace{2cm} &\hspace{0.0cm} Fit 1 (MeV) & \hspace{1.cm} Fit 2 (MeV)  & \hspace{1.cm} Fit 3 (MeV)  & \hspace{1.cm} Fit 4 (MeV)  \\
\hline
$O_1 = N_c \ \1 $                                                & $c_1 = 672 \pm 8$  & $c_1 = 673 \pm 7$      & $c_1 = 672 \pm 8$   & $c_1 = 673 \pm 7$   \\
$O_2 = \ell^i s^i$                	                         & $c_2 = 18 \pm 19$   & $c_2 = 17 \pm 18$    & $c_2 = 19 \pm 9$    & $c_2 = 20 \pm 9$ \\
$O_3 = \frac{1}{N_c}S^iS^i$                                      & $c_3 = 121 \pm 59$  & $c_3 = 115 \pm 46$   & $c_3 = 120 \pm 58$    & $c_3 = 112 \pm 42$\\
$O_4 = \frac{1}{N_c}\left[T^aT^a-\frac{1}{12}N_c(N_c+6)\right]$  & $c_4 = 202 \pm 41$  & $c_4 = 200\pm 40$   & $c_4 = 205\pm 27$   & $c_4 = 205\pm 27$  \\
$O_5 =  \frac{3}{N_c} L^{i} T^{a} G^{ia}$                  & $c_5 = 1 \pm 13$     &   $c_5 = 2 \pm 12$  &   \\ 
$O_6 =  \frac{15}{N_c} L^{(2)ij} G^{ia} G^{ja}$                  & $c_6 = 1 \pm 6$     &   &   $c_6 = 1 \pm 5$ \\ 
\hline
$B_1 = -\mathcal{S}$                                             & $d_1 = 108 \pm 93$  & $d_1 = 108 \pm 92$ & $d_1 = 109 \pm 93$  & $d_1 = 108 \pm 92$  \\     
\hline                  
$\chi_{\mathrm{dof}}^2$                                          &  $1.23$             & $0.93$   & $0.93$      & $0.75$\\
\hline \hline
\end{tabular}}

\end{table*}

When  hyperons are included in the analysis, the SU(3) symmetry must be broken and the mass operator takes the following general 
form \cite{JL95} 
\begin{equation}
\label{massoperator}
M = \sum_{i}c_i O_i + \sum_{i}d_i B_i .
\end{equation} 
The formula contains two types of operators. The first type are the operators $O_i$,  which are 
invariant under  SU($N_f$) and are defined as  
\begin{equation}\label{OLFS}
O_i = \frac{1}{N^{n-1}_c} O^{(k)}_{\ell} \cdot O^{(k)}_{SF},
\end{equation}
where  $O^{(k)}_{\ell}$ is a $k$-rank tensor in SO(3) and  $O^{(k)}_{SF}$
a $k$-rank tensor in SU(2)-spin.  Thus $O_i$ are rotational invariant.
For the ground state one has $k = 0$. The excited
states also require  $k = 1$ and $k = 2$ terms. 
The rank $k = 2$ tensor operator of SO(3) is
\begin{equation}\label{TENSOR} 
L^{(2)ij} = \frac{1}{2}\left\{L^i,L^j\right\}-\frac{1}{3}
\delta_{i,-j}\vec{L}\cdot\vec{L},
\end{equation}
which we choose to act on the orbital wave function $|\ell m_{\ell} \rangle$  
of the whole system of $N_c$ quarks (see  Ref. \cite{Matagne:2005gd} for the normalization 
of $L^{(2)ij}$). 
The second type are the operators  $B_i$ which are 
SU(3) breaking and are defined to have zero expectation values for non-strange baryons.
Due to the scarcity of data on the hyperons  here we consider only one hyperon and 
accordingly include only one of these operators,
namely $B_1 = - \mathcal{S}$ where $\mathcal{S}$ is the strangeness.

The values of the coefficients $c_i$ and $d_i$
which encode the QCD dynamics are determined from numerical fits to data.
Table  \ref{operators}  gives the list of  $O_i$  and $B_i$ operators together with  their coefficients,  which
 we believe to be the most relevant for the present study.  
The choice is based on our previous experience with the $N$ = 1 band \, \cite{Matagne:2011fr}.
In this table
the first nontrivial operator is the spin-orbit operator $O_2$.
In the spirit of the Hartree picture \cite{WITTEN}, generally adopted 
for the description of baryons,  we identify the 
spin-orbit operator with the single-particle operator 
\begin{equation}\label{spinorbit}
\ell \cdot s = \sum^{N_c}_{i=1} \ell(i) \cdot s(i),
\end{equation}
the matrix elements of which are of order $N^0_c$.
For simplicity  we ignore 
the two-body part of the spin-orbit operator, denoted by
$1/N_c\left(\ell \cdot S_c\right)$ in Ref. \cite{CCGL},
as being of a lower order  (there the lower case operators $\ell(i)$  act on
the excited quark and  $S_c$ is  the core spin operator). 
The analytic expression of the matrix elements of $O_2$ is given in the Appendix.

\begin{table*}[h!]
\begin{center}
\caption{Diagonal matrix elements of the operators $O_i$ for  the $[\bf{70}, \ell^-] $ multiplets  ( $\ell$ = 1, 2, 3)  of the 
$N$ = 3 band.}
\label{Matrix}
\renewcommand{\arraystretch}{2.3} {\scriptsize
\begin{tabular}{lccccccc}
\hline

\hline
  &\hspace{0cm}  $O_1$ &\hspace{0cm} $O_2$ & \hspace{0cm}$O_3$ &\hspace{0.2cm}&
  $O_4$ & \hspace{0cm}$O_5$     & \hspace{0cm}$O_6$ \\
  \hline
$^4N[{\bf 70},3^-]_{9/2}$ & $N_c$  & $\frac{3}{2}$  &  $\frac{15}{4N_c}$  && $\frac{3}{4N_c}$ 
&  $-\frac{9(N_c+3)}{4N_c}$   & $-\frac{75(N_c-1)}{8N_c}$   \\
$^2N[{\bf 70},3^-]_{7/2}$ & $N_c$  & $\frac{2N_c-3}{2N_c}$ & $\frac{3}{4N_c}$ && $\frac{3}{4N_c}$  
& $\frac{9}{2N_c}$  & $0$  \\
$^4N[{\bf 70},3^-]_{5/2}$ & $N_c$  & $-\frac{7}{6}$ &  $\frac{15}{4N_c}$  && $\frac{3}{4N_c}$ 
& $-\frac{7(N_c+3)}{4N_c}$   & $\frac{45(N_c-1)}{8N_c}$ \\
$^2N[{\bf 70},3^-]_{5/2}$ & $N_c$  & $-\frac{2(2N_c+3)}{3N_c}$ & $\frac{3}{4N_c}$ && $\frac{3}{4N_c}$
& $-\frac{6}{N_c}$   & $0$ \\
$^4N[{\bf 70},3^-]_{3/2}$ & $N_c$  & $-2$           &  $\frac{15}{4N_c}$   && $\frac{3}{4N_c}$
& $-\frac{3(N_c+3)}{N_c} $    & $-\frac{45(N_c-1)}{2N_c}$ \\
$^2N[{\bf 70'},1^-]_{3/2}$ & $N_c$ & $\frac{2N_c-3}{6N_c}$ & $\frac{3}{4N_c}$ && $\frac{3}{4N_c}$ & $ \frac{3}{2N_c} $   & $0$  \\
$^2N[{\bf 70'},1^-]_{1/2}$ & $N_c$ & $-\frac{2N_c-3}{3N_c}$ & $\frac{3}{4N_c}$ && $\frac{3}{4N_c}$  & $ -\frac{3}{N_c}$  & $0$ \\
$^2\Delta[{\bf 70},3^-]_{7/2}$ & $N_c$ & $-\frac{1}{2}$ & $\frac{3}{4N_c}$ && $\frac{15}{4N_c}$ 
& $\frac{9(N_c+1)}{4N_c}$ & $0$ \\
$^2\Delta[{\bf 70},2^-]_{5/2}$ & $N_c$ & $-\frac{1}{3}$ & $\frac{3}{4N_c}$ && $\frac{15}{4N_c}$ 
& $\frac{3(N_c+1)}{2N_c}$  & $0$  \\
$^2\Lambda[{\bf 70},3^-]_{7/2}$ & $N_c$ & $\frac{3}{2}$ & $\frac{3}{4N_c}$ && $-\frac{2N_c+3}{4N_c}$ 
& $-\frac{3(N_c-3)}{4N_c}$  & $0$  \\
\hline \hline
\end{tabular}}
\end{center}
\end{table*}

The spin operator $O_3$ and the flavor operator $O_4$ are two-body and linearly independent. 
The expectation values of $O_3$ are simply equal to $\frac{1}{N_c} S ( S + 1 )$ where $S$ is the spin 
of the whole system. They are given in Table \ref{Matrix}.

Note that the definition of the operator $O_4$,  indicated in Table \ref{operators},
 is such as to recover the matrix elements of
the usual $1/N_c (T^a T^a)$  in SU(4) by subtracting the quantity  $(N_c+6)/12$.
This is understood  by using Eq. (30) of Ref. \cite{Matagne:2008kb} for the matrix 
elements of $1/N_c (T^a T^a)$ extended to SU(6).  Then, as one can see from Table
\ref{Matrix} it turns out that  
the expectation values  of  $O_4$   are positive for octets and decuplets 
 and of order  $N^{-1}_c$, as in SU(4), and negative   and of order $N^0_c$  for flavor singlets 
(see the Appendix for details).

The operators  $O_5$  and $O_6$ are also two-body, which means that they carry a 
factor $1/N_c$  in the definition.  However,  as $G^{ia}$
sums coherently, it introduces an extra factor $N_c$ and makes the matrix 
elements of $O_5$ and $O_6$ of order $N^0_c$, as seen from Table    \ref{Matrix}. 
These matrix elements are obtained from the formulas (B2) and (B4) of Ref. \cite{Matagne:2011fr}
where the multiplet $[70,1^-]$ has been discussed.  Interestingly,  when $N_c$ = 3, 
the contribution of $O_5$ cancels out for flavor singlets,   like 
for $\ell$ = 1 \cite{Matagne:2011fr}. This property follows from the analytic form of the isoscalar factors 
given in Ref. \cite{Matagne:2011fr}.

We remind that 
the SU(6) generators $S^i$, $T^a$ and $G^{ia}$  and the $O(3)$ generators $L^i$ of Eq. (\ref{TENSOR}) act on the
total wave function of the $N_c$ system of quarks as proposed in Refs.  \cite{Matagne:2006dj}, 
\cite{Matagne:2008kb} and  \cite{Matagne:2011fr}.  The advantage of this procedure over the standard one, where the system is 
separated into a ground state core + an excited quark \cite{CCGL},
is that the number of relevant operators needed in the fit is usually smaller than the number of data and 
it allows a better understanding of their role in the mass formula, in  particular the role of the isospin operator $O_4$ 
which has always been omitted in the symmetric core + excited quark procedure. 
We should also mention that in our  approach the permutation symmetry is exact \cite{Matagne:2006dj}.

Among the operators containing angular momentum components, besides the spin-orbit, we have included the operators 
$O_5$ and $O_6$, to check whether or not  they bring insignificant  
contributions,  as it was  in the $N$ = 1 band.  From  Table \ref{operators}
one can see that their coefficients are indeed negligible either included together as in Fit 1 or separately as in Fit 2 and 3.




\begin{table*}[h!]
\begin{center}
\caption{Partial contributions and the total mass (MeV) predicted by the $1/N_c$ expansion method obtained from Fit 4. The last two columns indicate the empirically
known masses and the resonance name and status (whenever known). }\label{MASSES1}
\renewcommand{\arraystretch}{2.5}{\scriptsize
\begin{tabular}{lrrrrrrrrr}\hline \hline
                    &      \multicolumn{5}{c}{Part. contrib. (MeV)}  & \hspace{0.75cm} Total (MeV)   & \hspace{0.75cm}  Exp. (MeV)\hspace{0.75cm}& &\hspace{0.75cm}  Name, status \hspace{.0cm} \\

\cline{2-6}
                    &   \hspace{.35cm}   $c_1O_1$  & \hspace{.35cm}  $c_2O_2$ & \hspace{.35cm}$c_3O_3$ &\hspace{.35cm}  $c_4O_4$   &  \hspace{.35cm}  $d_1B_1$   &      \\
\hline
$^4N[{\bf 70},3^-]_{9/2}$      & 2018 & 29 & 140 & 51 &  0  &$2238\pm 46$  & $2275\pm 75$ & & $G_{19}(2250)$****  \\
$^2N[{\bf 70},3^-]_{7/2}$      & 2018 & 10 &  28 & 51 &  0  &$2107\pm 17$  & $2150\pm 50$ & & $G_{17}(2190)$**** \\
$^4N[{\bf 70},3^-]_{5/2}$      & 2018 &-23 & 140 & 51 &  0  &$2186\pm 41$  & $2180\pm 80$ & & $D_{15}(2200)$**\\
$^2N[{\bf 70},3^-]_{5/2}$      & 2018 &-39 &  28 & 51 &  0  &$2058\pm 14$  & $2060\pm 15$ & & $D_{15}(2060)$\\
$^4N[{\bf 70},3^-]_{3/2}$      & 2018 &-39 & 140 & 51 &  0  &$2170\pm 42$  & $2150\pm 60$ & & $D_{13}(2150)$\\
$^2N[{\bf 70'},1^-]_{3/2}$     & 2018 &	 3 &  28 & 51 &  0  &$2101\pm 14$  & $2081\pm 20$ & & $D_{13}(2080)$* \\
$^2N[{\bf 70'},1^-]_{1/2}$     & 2018 & -7 &  28 & 51 &  0  &$2091\pm 12$  & $2100\pm 20$ & & $S_{11}(2090)$*\\
\hline
$^2\Delta[{\bf 70},3^-]_{7/2}$ &  2018 & -10  & 28 & 256 &  0  &  $2292\pm 25$  & $2200\pm 80$ & & $G_{37}(2220)$* \\
$^2\Delta[{\bf 70},2^-]_{5/2}$ &  2018 & -7  & 28 & 256 &  0  &  $2295\pm 25$  & $2305\pm 26$ & & $D_{35}(2350)$* \\ 
\hline
$^2\Lambda[{\bf 70},3^-]_{7/2}$& 2018 &29 &  28 &-153&108  &$2030\pm 82$  & $2030\pm 82$ & & $G_{07}(2100)$**** \\ 
\hline \hline
\end{tabular}}
\end{center}
\end{table*}

\section{Results and discussion}
We have performed four distinct numerical fits of the mass formula  (\ref{massoperator}) to the experimental data.
The corresponding dynamical coefficients 
$c_i$ and $d_i$ together with the values of $\chi^2_{dof}$ are    listed in Table \ref{operators}. 
Fit 1 is made with all operators. Fit 2 and Fit 3  are made by removing one operator  and Fit 4 is made only the first 
four operators.  
It turns out that  the contributions of angular dependent operators $O_5$ and $O_6$  are negligible
but that of the spin-orbit operator, which is quite small, remains  important. 
The values of $\chi^2_{dof}$ are good and the error bars of the coefficients
suggest that the choice of the operators we have made provides a reliable fit.

As already mentioned, in the $1.9 - 2.4$ GeV region the experimental data are rather scarce. 
For this reason,  besides the  eight  resonances provided  by  the Particle Data Group (PDG) \cite{PDG} 
we have also included  two more,  proposed in Refs.  \cite{Anisovich:2011ye,Anisovich:2011fc}. 
They do not have a status yet.  Therefore, in all,  we have included ten resonances in the fit from which one is a hyperon.

The experimental masses of four star resonances are from
the Summary Table of PDG. 
For the two star resonance $D_{15}(2200)$
we took the mass  indicated in  the  Baryon Particle Listings of PDG  
as due to Cutkosky et al. \cite{Cutkosky:1980rh} and for the one star resonance 
 $D_{13}(2080)$ the mass  due to Hoehler et al. \cite{hoehler79}.
The experimental mass of $S_{11}(2090)$ was taken 
as the average of the masses  obtained by Hoehler et al. and Cutkosky et al. in the partial wave analysis of the $\pi N$ scattering.

The  baryon masses obtained from the mass formula  (\ref{massoperator}) with the 
coefficients  from Fit 4 and matrix elements from Table \ref{Matrix}
are presented in Table \ref{MASSES1}. Partial contributions of different operators to the total mass are also indicated.   
One can see that the spin operator $O_3$ brings a dominant contribution to the splitting in $^4N[{\bf 70},3^-]$ states 
and the isospin operator is as important, or even more, in the $\Delta$ and the $\Lambda$ resonances. In the latter,
the negative sign of $O_4$ matrix elements helps in lowering the calculated mass close down to the experimental value.

The first observation regarding the multiplet structure is related to the the $G_{17}(2190)$ four star resonance. 
To obtain a good fit we had to 
interpret it as  a    $^2N[{\bf 70},3^-]$  state.  If so,  Table \ref{MASSES1}
implies that it  forms a doublet  with the newly suggested 
$D_{15}(2060)$ resonance of Refs.  \cite{Anisovich:2011ye,Anisovich:2011fc}. 
If  this resonance is the same
as the newly suggested $N(2070)D_{15}$ resonance of Ref. \cite{Bartholomy:2007}
we should be  in agreement with the  latter authors, who proposed a doublet.

As a matter of fact we have found out that the inclusion of the resonance $S_{11}(2090)$, interpreted as a radial
excitation belonging to the $[{\bf 70'},1^-]$ multiplet, improves the fit, which gives confidence in this interpretation.
This resonance appears therefore as the spin-orbit partner of $D_{13}(2080)$.

Regarding the $\Delta$ resonances, our analysis shows that  we have interpreted  $D_{35}(2350)$ as a member of a 
$[{\bf 70},2^-]$ multiplet. This is  inspired by  
the quark model results of Ref. \cite{Stancu:1991cz} where this multiplet, having $\ell$ = 2,  is the highest in the spectrum 
of the spin-independent Hamiltonian with a relativistic kinetic energy and a linear confinement. Such a high value 
is expected to lead to  a  mass as large as that  of the mentioned resonance.  This choice suggests a kind
of agreement between quark models and the present fit, well in the spirit of Ref. \cite{Semay:2007cv}.

The value of the coefficient $c_1$ found in our best fit $c_1 = 673 \pm 7$ MeV is  smaller than  $c_1 = 731 \pm 17$ MeV of Ref. 
\cite{Goity:2007sc} but not far from the estimate  $c_1 \approx 640$ MeV
which can be extracted from Fig. 1  of Ref. \cite{Matagne:2005gd}
where a linear  dependence of $c_1$ on  the excitation energy, or alternatively on the band number $N$,  was found. 
Note that such an energy dependence is reproduced by the formula (29) of Ref. \cite{Semay:2007cv}
where the compatibility between the $1/N_c$ expansion method and semi-relativistic quark  models with a linear
confinement  was discussed.  This compatibility is confirmed by the present results.
The value of $c_2 = 20 \pm 9$ MeV is  practically identical  with that obtainable from Fig. 1 
of Ref. \cite{Matagne:2005gd}. 

From the comparison of our results with the  ``new'' resonances reported in Refs. \cite{Anisovich:2011ye,Anisovich:2011fc} 
we can make the following comments. We do not support the interpretation of the $D_{15}(2060)$ resonance as a member 
of the quartet  (\ref{quartet}),
inasmuch as we interpret this resonance as a member of the doublet $^2N[{\bf 70},3^-]$.  But   
$D_{13}(2150)$ is a  member of the quartet $^4N[{\bf 70},3^-]$ together with $G_{19}(2250)$ and $D_{15}(2200)$.  If this 
interpretation is valid it remains to find the $J$ = 7/2  member,  not observed yet. 
 
\section{Conclusions}

Using the $1/N_c$ expansion method 
we have analyzed the multiplet structure of high-lying negative parity resonances, located in the 1900 MeV -  2400 MeV region,
 supposed to belong to the $N$ = 3 band in the SU(6) $\times$ O(3) classification. 
 Our results are largely consistent with the recent experimental analysis  of 
Refs. \cite{Bartholomy:2007,Anisovich:2011ye,Anisovich:2011fc}.  A possible future observation of a $7/2^-$
resonance would be of great help in understanding the $^4N[{\bf 70},3^-]$ multiplet and the structure of the
$N$ = 3 band in general. 

The simplified method we have used allows us to include a small number of terms in the mass formula and 
easily identify the most dominant operators 
to order $\mathcal{O}(N^{-1}_c)$. 
As a common feature with the SU(4) and SU(6) analysis of the $N$ = 1 band,
we found  that the isospin operator $O_4$, neglected in the standard core + excited quark approach,
contributes to the mass of $\Delta$'s 
with a coefficient $c_4$ with a magnitude comparable to that  of the coefficient $c_3$ of the spin operator $O_3$ in $N^*$ resonances. 
In addition the role of the operator $O_4$ is crucial in describing  the flavor singlet four star resonance $\Lambda(2100)G_{07}$
included in the fit. 

Future discoveries will help to improve our study and confirm or infirm the present  multiplet interpretation.

\appendix

\section{}

We remind  that
the matrix elements  of the spin-orbit operator $O_2$ between states with spins $S$ and $S'$ are given by 
\begin{eqnarray}
\lefteqn{\langle \ell' S'J'J'_3;I'I'_3 | \ell \cdot s | \ell S JJ_3; I I_3 \rangle_{p=2}  = }\nonumber \\& & ( -1)^{J+\ell+1/2} \delta_{J'J} \delta_{J'_3J_3}\delta_{\ell'\ell} \delta_{I'I} \delta_{I'_3I_3} \sqrt{\frac{3}{2}(2 S + 1) (2 S' + 1) \ell (\ell + 1)(2 \ell + 1)} 
 \left\{\begin{array}{ccc}
        \ell & \ell & 1 \\
	S & S' & J
      \end{array}\right\} 
\nonumber \\& &\times \sum_{p_1,p_2} ( -1)^{-S_c} c^{[N_c-1,1]}_{p_1p_2}(S') c^{[N_c-1,1]}_{p_1p_2}(S)
 \left\{\begin{array}{ccc}
        1 & \frac{1}{2} &  \frac{1}{2}\\
	S_c & S & S'
      \end{array}\right\}.  
\end{eqnarray}
The quantities $c^{[N_c-1,1]}_{p_1p_2}(S)$, are  a short hand notation for the isoscalar factors of the permutation group
of $N_c$ quarks,   denoted by $K([f']p'[f'']p''|[f]p)$ in Ref.  \cite{Matagne:2008fw} (see also Ref. \cite{Stancu:1991rc}). 
In that notation 
 $p$, $p_1$ and $p_2$ represent the position of the $N_c$-th quark in the spin-flavor, spin and flavor 
parts of the wave function, of partitions $[f]$, $[f']$ and $[f'']$ respectively. 
Actually $c^{[N_c-1,1]}_{p_1p_2}(S)$ are functions of the spin $S$ and the number $N_c$ of quarks.
By definition, the core + excited quark wave function  \cite{CCGL} has  $p =2$.
This expression is equivalent to  Eq. (A7) of Ref. \cite{CCGL}.  The correspondence in the isoscalar factors denoted there by $c_{\rho \eta}$ is
\begin{equation}\label{coresp}
       c^{[N_c-1,1]}_{1 1} (S) \rightarrow c_{0-}; ~~~~c^{[N_c-1,1]}_{2 2} (S) \rightarrow c_{0+};
~~~ c^{[N_c-1,1]}_{1 2} (S) \rightarrow c_{++}; ~~~~c^{[N_c-1,1]}_{2 1} (S) \rightarrow c_{--}.
\end{equation}

We also remind that the matrix elements of the isospin operator $O_4$ as defined in Table \ref{operators}
requires the knowledge of the expectation value of the SU(3) Casimir operator $T^a T^a$.  Labelling the flavor states by $(\lambda,\mu)$
this is 
\begin{equation}\label{casimir}
\langle T^a T^a \rangle = \frac{1}{3} (\lambda^2 + \mu^2 + \lambda \mu + 3 \lambda + 3 \mu)
\end{equation}
For the flavor states of $N_c$ quarks we have
\begin{itemize}
\item
$^28 ~~ or ^48 : ~~~  \lambda = 1,  ~~~\mu = \frac{N_c - 1}{2}$, ~~~ $\langle T^a T^a \rangle = \frac{(N_c + 3)^2}{12}$,
\item
$^210: ~~~  \lambda = 3,  ~~~\mu = \frac{N_c - 3}{2}$, ~~~ $\langle T^a T^a \rangle = \frac{N^2_c + 6 N_c + 45}{12}$,
\item
$^21: ~~~  \lambda = 0,  ~~~\mu = \frac{N_c - 3}{2}$, ~~~ $\langle T^a T^a \rangle = \frac{N_c (N_c  + 6) - 3 (2 N_c + 3)}{12}$,
\end {itemize} 
The last case has been discussed in Ref.  \cite{Matagne:2008kb}. Accordingly the expectation value  of $O_4$  is 
negative for flavor singlets.

The analytic expressions of the matrix elements of $O_5$  and $O_6$ can be found in Ref.  \cite{Matagne:2011fr}
together with the corresponding isoscalar factors of the SU(3) generators which we do not reproduce here. 

\vspace{2cm}
 
{\bf Acknowledgments}
 The work of one of us (N. M.) was supported by F.R.S.-FNRS (Belgium).



\end{document}